\documentclass{edp-conf}
\usepackage{graphicx,psfig, epsf,epsfig}
\begin{document}

\TitreGlobal{SF2A 2004}

\title{Simultaneous INTEGRAL/RXTE observations of GRS 1915+105}
\author{Rodriguez, J.}\address{CEA Saclay, DSM/DAPNIA/SAP, 91191 Gif Sur Yvette, \& ISDC Versoix, Switzerland}
\author{Hannikainen, D.C.}\address{Observatory, University of Helsinki, Finland}
\author{Shaw, S.E.}\address{University of Southampton, UK \& ISDC Versoix, Switzerland}
\author{Cabanac, C.}\address{LAOG, Grenoble, France}
\author{Fuchs, Y.}\address{CEA Saclay, France}
\runningtitle{Simultaneous INTEGRAL/RXTE observations of GRS 1915+105}
\setcounter{page}{1}
\index{Rodriguez, J.}
\index{Hannikainen, D.C.}
\index{Shaw S.}
\index{Cabanac, C.}
\index{Fuchs, Y.}

\maketitle
\begin{abstract}We report here the first results of the spectral and timing analysis of
our simultaneous {\it INTEGRAL/RXTE} observations of GRS 1915+105.The 
first observation ever performed with {\it INTEGRAL} revealed a new class of variability,
where changes of luminosity seem driven by changes in the Comptonising medium. 
The spectro-temporal study of our data taken later, in the steady state, could show the 
influence of the compact jet in the hard X-rays.
 \end{abstract}
%
\section{Introduction}
\begin{figure}[htbp]
\centering
\epsfig{file=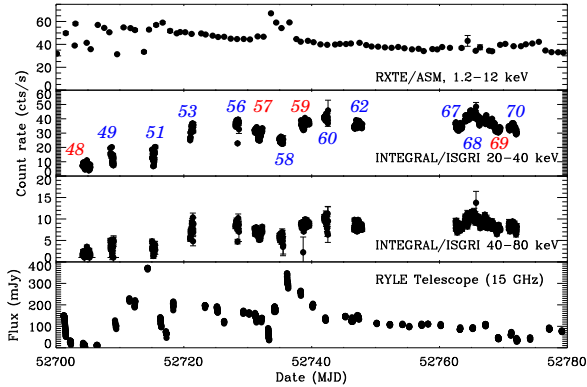,width=8cm}
\caption{\scriptsize{RXTE/ASM (top), INTEGRAL/ISGRI 20-40 and 40-80 keV (middle), and Ryle 15 GHz (bottom)
light curves of GRS~1915+105 covering March--May 2003. The INTEGRAL revolution numbers are 
indicated. Light grey numbers correspond to the dates where we had simultaneous {\it RXTE} pointings.}}
\label{fig:lc}
\end{figure}

 GRS 1915+105 is one of the most fascinating X-ray sources in our Galaxy. It is the biggest
stellar mass black hole (BH) with a mass of 14$\pm$4.4 M$_\odot$ (Harlaftis and Greiner 2004),
one of the brightest X-ray sources, and a definite source of apparent superluminal ejection 
(Mirabel \& Rodr\'{\i}guez 1994). An up-to-date accurate review of the source can be found 
in Fender \& Belloni (2004). \\
\indent Since the launch of the {\it INTEGRAL} observatory we have monitored the source with 
dedicated simultaneous {\it INTEGRAL} and {\it RXTE} pointed observations but also, most of the time,
with  (nearly) simultaneous observations from the ground in the radio and/or infrared 
domain.
While our second year of monitoring is about to start, we report here the 
results of the first  year of this programme focusing mainly on the data acquired in March-May 2003
during an unprecedented high energy coverage (Fig. \ref{fig:lc}) 

\section{First {\it INTEGRAL} observation (revolution 48): a new class of variability}
\begin{figure}[htbp]
\centering
\epsfig{file=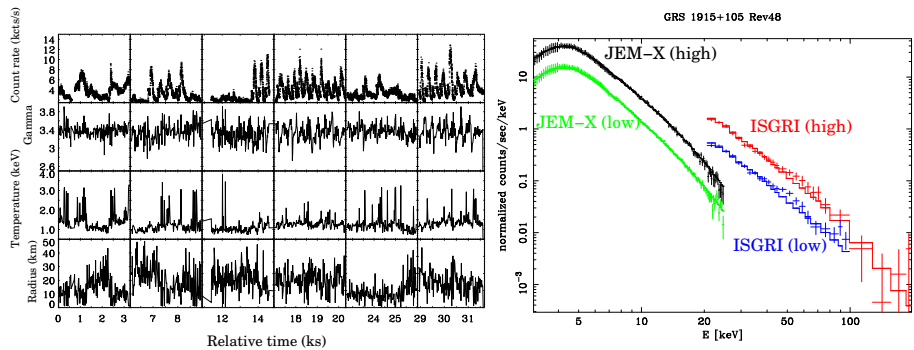,width=12cm}
\caption{\scriptsize{{\bf Left:} Light curve and spectral parameters obtained from fits 
to 16 s PCA spectra. {\bf Right:} Average spectra obtained with INTEGRAL from high and low luminosity states
of the whole 100 ks observation.}}
\label{fig:rev48spec}
\end{figure}

The source is extremely variable in both the soft and hard X-rays (Hannikainen et al.
2003, top left panel of Fig. \ref{fig:rev48spec}), while the average spectrum is very soft. 
Our preliminary analysis showed that the source was in a class of variability never observed before 
(Hannikainen et al. 2003) showing high spikes with quasi periodic period of about 5 minutes,
hard rises and soft declines. While a first spectral analysis of an average spectrum was given in 
Hannikainen et al. (2003), the variability implies a better definition 
of the good time interval from which to accumulate spectra if one wants to understand the origin of the variations. With RXTE/PCA 
we produced spectra every 16~s and fitted them with a simple model of multi colour disc black body
and a power law, while we accumulated spectra from INTEGRAL/JEM-X and ISGRI from the high level 
and the low level (Fig. \ref{fig:rev48spec}, Hannikainen et al. 2005, 
in prep.). The best fit parameters obtained from the INTEGRAL spectra are kT$_{in}$=1.42$\pm$0.03 keV and
$\Gamma=3.46\pm0.05$ for the high luminosity periods, and  kT$_{in}=1.28\pm0.02$ keV and $\Gamma=3.34\pm0.05$
for the low luminosity one.
While we can see a slight evolution of the disc temperature and the powerlaw photon index, the 
inner radius of the disc stays approximatively the same. The flux of the power law however changes drastically.
It seems, therefore, that the variability of this class, unlike most of the others, is driven by changes
in the powerlaw component usually attributed to Comptonisation of the disc photons by a ``corona''. This is 
compatible with the PCA analysis (Fig. \ref{fig:rev48spec}) where no apparent correlation between 
the disc parameters and the spikes are visible, while the variations of the powerlaw photon index may be 
tightly linked to changes of luminosity. Analysis of this observation with more physical elaborated models 
is in progress (Hannikainen et al. 2005, in prep.).

\section{Steady state observation: Energy dependence of the low frequency QPOs}
From {\it INTEGRAL} revolution 53 on, GRS 1915+105 was found in its steady ``hard'' state (class $\chi$ from Belloni 
et al. 2000). The huge multi-wavelength campaigns performed during revolutions 57 and 62 (Fuchs et al. 2003, 2005, 
in prep.) revealed the simultaneous presence of a strong compact jet, a hard X-ray tail, and strong 
Quasi Periodic Oscillation (QPO, at least 
during revolution 57 when we had simultaneous {\it RXTE} observations), while monitoring with the Ryle Telescope
showed that the source was very active in radio at the same time (Fig. \ref{fig:lc}).
The analysis of our whole {\it RXTE} data set revealed the presence of strong low frequency QPOs (LFQPO) in each observation
with frequencies between 1-2.5 Hz, and amplitudes between 11.5-15 \% rms (Rodriguez et al. 2004). 
\begin{figure}[htbp]
\centering
\epsfig{file=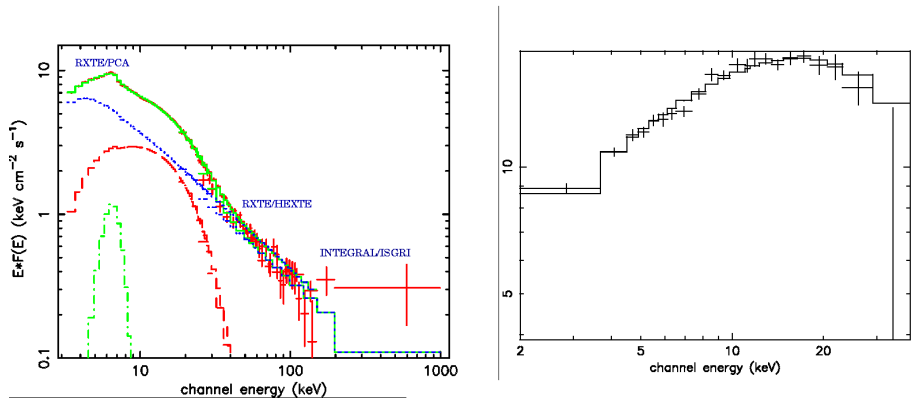,width=12cm}
\caption{\scriptsize{{\bf Left:} Simultaneous {\it INTEGRAL/RXTE} fit of the data from revolution 57, with 
a model consisting of comptt and an extra powerlaw to account for excess at high energy. {\bf Right:} ``QPO spectrum''
plus best fit model of a cut off powerlaw.}}
\label{fig:qpospec}
\end{figure}
Unlike usually observed no correlation was found between the source count rate and the QPO frequency.
We studied the energy dependence of the QPO amplitude (QPO spectra) in 22 spectral bins. This allowed us to perform
spectral fits and we showed that although the source spectral parameters were similar in all {\it RXTE} observations,
the QPO spectra differed markedly (Rodriguez et al. 2004): while in the first observation (Fig. \ref{fig:qpospec}, right) 
an obvious cut-off was found (at $\sim22$ keV), in the following ones the cut-off energy  either evolved towards
higher energy (up to $\sim 30$ keV), or simply disappeared.\\
\indent Our spectral studies revealed that the spectra of GRS 1915+105 could be well represented by a thermal 
comptonisation component, and an extra powerlaw accounting for emission at higher energies (Fig. \ref{fig:qpospec}, left).
\indent While quite puzzling, those results, and especially the evolution of the cut-off energy in the QPO spectra
can be easily understood if we assume that part of the hard X-rays are emitted by the compact jet (through
direct synchrotron or synchrotron self-Compton as expected see e.g. Markoff \& Nowak 2004), while another part would come 
from inverse Comptonisation on the basis of the jet. In this context the QPO would originate in the latter component, 
while the jet emission would not contain any modulation. The QPO spectra would therefore be  due to the relative 
contribution of both components to the total X-ray spectrum: the stronger the jet, the lower the cut-off energy
of the QPO, and vice-versa.\\
\indent We point out that this interpretation qualitatively matches several observational/theoretical facts. 
The compact jet model has succesfully been used in the fitting of different BH binaries (e.g. Markoff et al. 2001). 
In GRS 1915+105, there is a need for an additional parameter to account for the hardest part of the spectrum. The compact
jet is detected (Fuchs et al. 2003), and is strongest when the QPO cut-off energy is the lowest (Rodriguez et al. 2004).
The QPO spectral shape seems to be qualitatively the same as the relative contribution of the Comptonised component 
to the overall spectrum (Fig. \ref{fig:qpospec}).\\


\end{document}